\documentclass[11pt]{elsarticle}

\pdfoutput=1

\makeatletter
\def\ps@pprintTitle{%
  \let\@oddhead\@empty
  \let\@evenhead\@empty
  \let\@oddfoot\@empty
  \let\@evenfoot\@oddfoot
}
\makeatother

\usepackage{url}
\usepackage{breakurl}
\usepackage[breaklinks,
            colorlinks = true,
            linkcolor = blue,
            urlcolor  = blue,
            citecolor = blue,
            anchorcolor = blue]{hyperref}

\usepackage{lineno}

\usepackage{graphicx}
\usepackage[margin=1.25in]{geometry}
\usepackage[usenames,dvipsnames]{color}

\usepackage{fancyhdr}
\fancypagestyle{plain}{%
  \fancyhf{}%
  \fancyhead[C]{}
  \fancyfoot[C]{\thepage}
}

\fancypagestyle{empty}{%
  \fancyhf{}%
  \fancyhead[C]{{\it HEP in developing or emerging regions for the online Particle Data Book}}
  \fancyfoot[C]{\thepage}
}
\begin{document}

\begin{frontmatter}


\title{High Energy Physics in Africa, Latin America and other developing regions}

\author[add1]{K\'et\'evi A. Assamagan\corref{cor1}}
\ead{ketevi@bnl.gov}
\author[add2]{Johan Sebastian Bonilla}
\author[add3]{Claudio Dib}
\author[add4]{Azwinndini Muronga}
\author[add5]{Heath B. O'Connell}
\author[add6]{Rogerio Rosenfeld}
\author[add7]{Suyog Shrestha}

\cortext[cor1]{Corresponding Author}

\address[add1]{Brookhaven National Laboratory, Physics Department, Upton, New York, USA}
\address[add2]{University of California, Davis, USA}
\address[add3]{Dept. of Physics and CCTVal, Universidad T\'ecnica Federico Santa Maria
Valparaiso, Chile}
\address[add4]{Faculty of Science, Nelson Mandela University, Gqeberha, South Africa}
\address[add5]{Fermi National Accelerator Laboratory, USA}
\address[add6]{Instituto de Física Teórica, UNESP and ICTP-SAIFR, Sao Paulo, Brazil}
\address[add7]{Washington College, Chestertown, MD USA}

\begin{abstract}
\noindent We summarize the current status of high energy physics (HEP) in Africa, Latin America, and other developing regions. 
\end{abstract}


\end{frontmatter}


%


\newpage

\section{Introduction}
\label{sec:why}
\noindent 
We attempt to describe HEP activities and efforts in Africa, Latin America and other regions that may be classified as developing or emerging. The report is not exhaustive and materials shown are based on the expert knowledge of the authors at the time of information gathering. It builds upon prior work done in the context of US particle physics prioritization exercise as detailed in Ref.~\cite{assamagan2022us} where the reader may find additional useful information.  

The narrative was developed in July 2023 and would require periodic updates, as the HEP landscape evolves and changes across the world. 

\section{HEP in Africa}
\label{sec:current}
\begin{figure}[!htpb]
\begin{center}
\includegraphics[width=\textwidth]{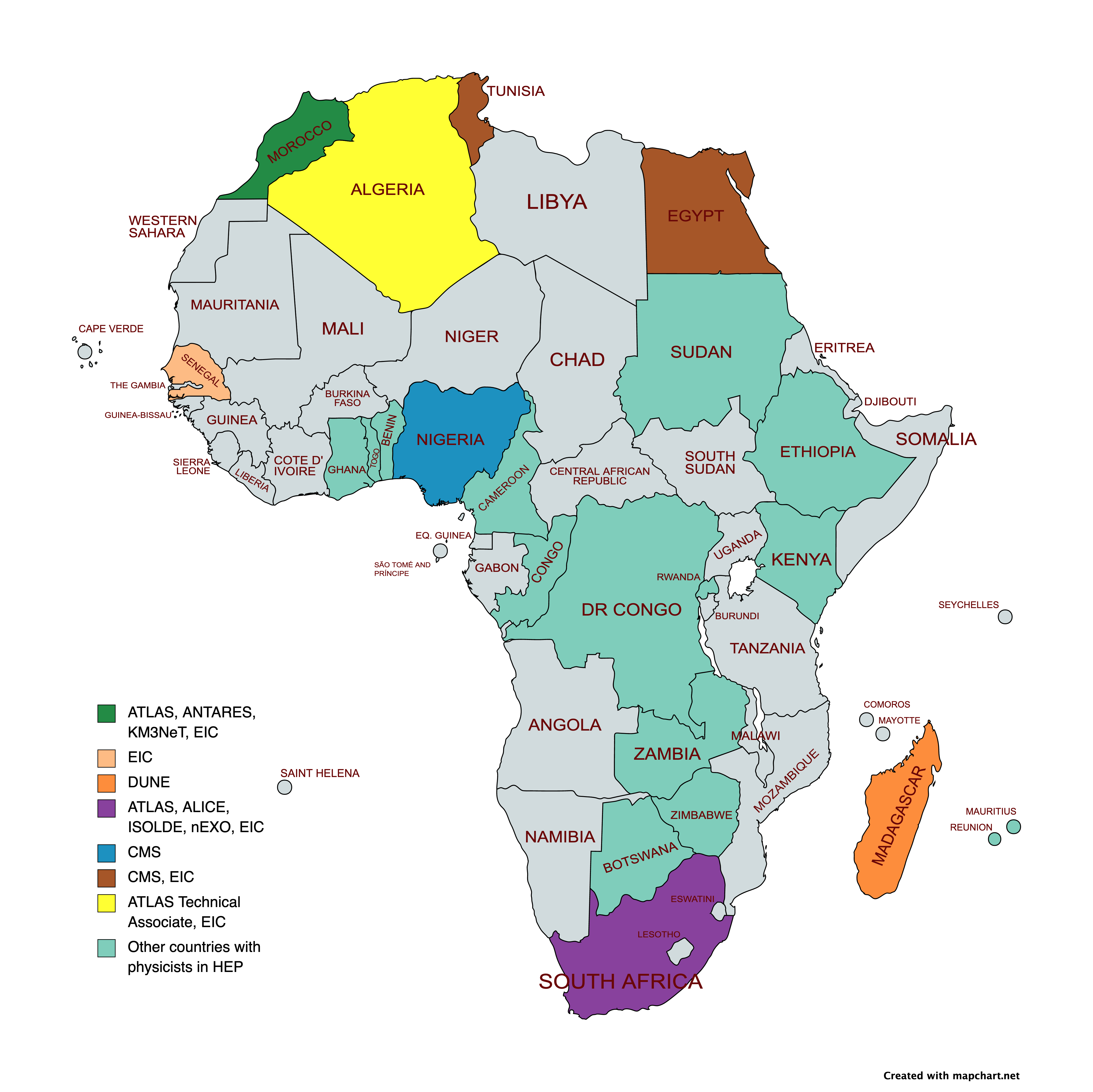}
\end{center}
\caption{African countries with formal programs or citizen in high energy physics.}
\label{fig:hep-africa}
\end{figure}

Figure~\ref{fig:hep-africa} shows the African countries with HEP physics programs. A handful of African countries---Morocco~\cite{atlas}, Egypt~\cite{Egypt-CERN} and South Africa~\cite{atlas, alice, isolde}---have HEP programs in theory and experiments at the LHC as described in Refs.~\cite{assamagan2022us, amhis2022high}. Morocco has been involved in the neutrino astrophysics experiments of ANTARES~\cite{Antares} and KM3Net~\cite{KM3Net}. The South African -- CERN program is managed at iThemba LABS, a nuclear and high energy physics research and education facility~\cite{iThemba}. South Africa has had a strong participation in JINR~\cite{SA-JINR} and is a member of nEXO, a neutrinoless double beta decay experiment~\cite{nEXo}. In 2016, Madagascar joined the DUNE Collaboration~\cite{dune}; more recently, Nigeria and Tunisia joined the CMS Collaboration~\cite{assamagan2022us, cms}; Algeria has become technical associate institute in ATLAS~\cite{atlas}. Senegal and South Africa have joined the EIC Collaboration~\cite{eic}. Capacity development in HEP are also organized through the African School of Physics~\cite{ASP}, the East African Institute for Fundamental Research~\cite{EAIFR}, and African Institute for Mathematical Sciences~\cite{AIMS}. Figure~\ref{fig:hep-africa} also shows countries with citizens involved in HEP, but the countries do not have organized and supported HEP activities; these citizens in HEP are employed elsewhere~\cite{cern-map} in Asia, Europe, North America, and to a lesser extend South Africa. 

The African physics community is currently in the process of developing the African strategy on fundamental and applied physics (ASFAP)~\cite{ASFAP} through a grassroots-led planning exercise. The collective efforts of the African HEP community will be captured in a final ASFAP report at the end of the strategy planning exercise. 

\section{HEP in Latin America}
\label{sec:LA}
The HEP community in Latin America (or more generally High Energy, Cosmology and Astroparticle Physics - HECAP)
initiated in 2018 its first Strategic Planning with a mandate from the Iberoamerican Science and Technology Ministries. The Snowmass model was used and a call for White Papers issued. With the 39 submitted White Papers by December 2019, three main documents were produced:\\
$\bullet$ a 85-pages Physics Briefing Book describing the activities in HECAP in Latin America;\\
$\bullet$ a Strategy Document containing 10 recommendations for the area;\\
$\bullet$ an Endorsement Letter from the High-Level Strategy Group.
All white papers and documents are publicly available at
the site of the Latin American Strategy Forum for Research Infrastructure (LASF4RI)~\cite{lasf4ripage}. The Physics Briefing Booked was also posted in the arXiv~\cite{LASF4RI2021} and the recommendations were recognized in the 2020 
Iberoamerican Science and Technology Ministerial Meeting~\cite{MinisterialDeclarationLASF4RI}.

\begin{figure}[!htpb]
\begin{center}
\includegraphics[width=\textwidth]{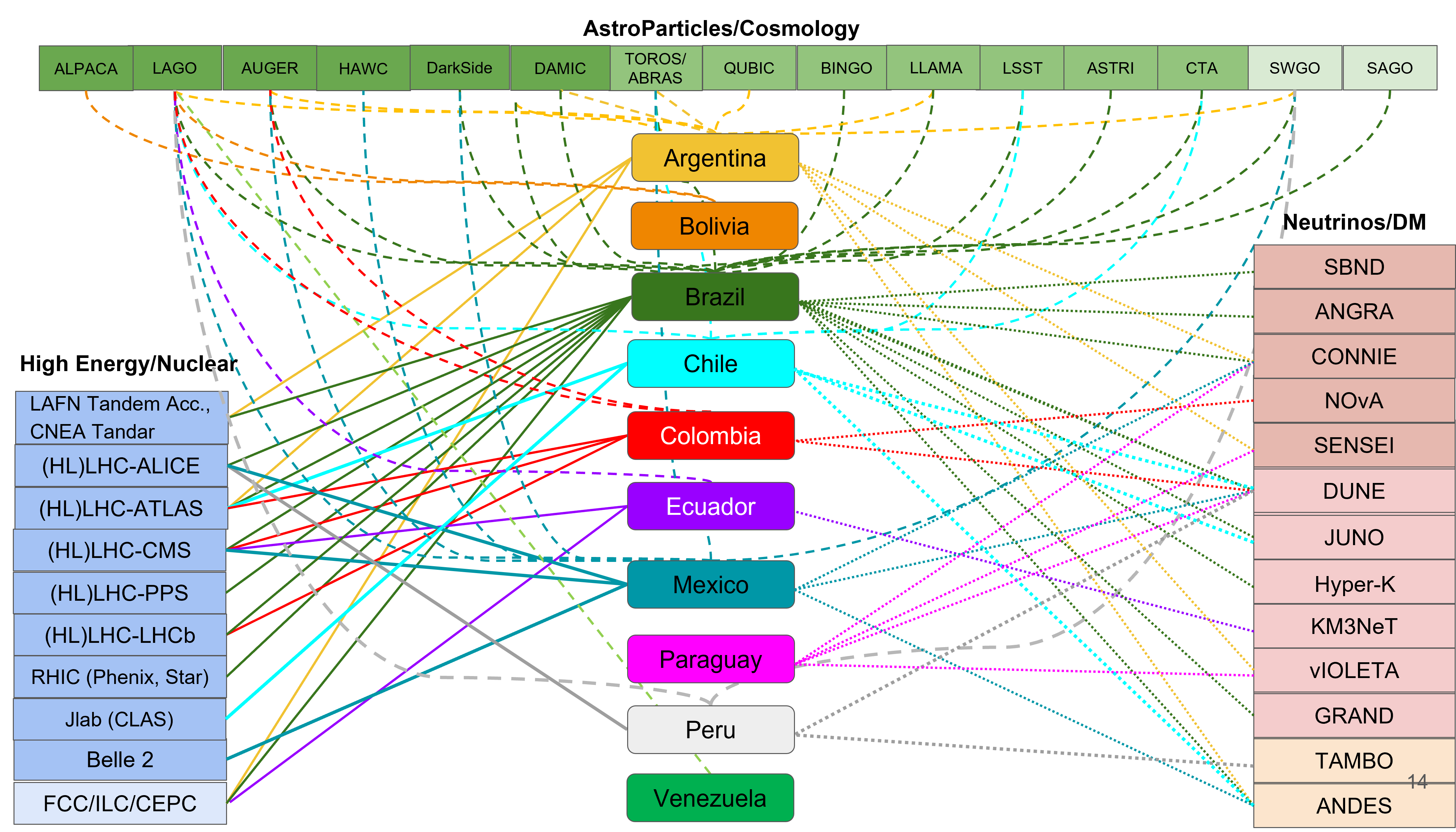}
\end{center}
\caption{Landscape reflecting the participation of the LA countries in different HECAP projects.}
\label{fig:hep-la}
\end{figure}

A complex landscape of LA participation in different international collaborations emerged from this effort and
is depicted in Figure~\ref{fig:hep-la}. In 2021, following the recommendations for the community to put in place 
a more robust structure, the Latin American Association for High Energy, Cosmology and Astroparticle Physics (LAA-HECAP) was
created and has more than 450 members and is hosted by the ICTP-South American Institute for Fundamental Research (ICTP-SAIFR).
ICTP-SAIFR started its activities in 2012 and hosts several events in HECAP every year, including workshops related to LASF4RI.
LAA-HECAP is seeking partnerships and funding for LA activities in HECAP, in addition to
coordinate the organization of the Latin American Symposium on High Energy Physics, the venue for scientific interaction in the community, which will have its XV edition in 2024 in Mexico. LASF4RI plans to start an update of the HECAP strategy in 2023, calling for Letters of Interest and White Papers by July 2024.

While large HEP experimental facilities today are used by worldwide collaborations, the existence of such facilities in a specific region does imply local development. So far, Latin America counts with large facilities in the field of Astroparticle Physics, with significant involvement of the LA scientific community: there is the Pierre Auger Observatory in Argentina~\cite{Augerpage}, the High Altitude Water Cherenkov observatory (HAWC) in Mexico~\cite{HAWCpage}, the Cherenkov Telescope Array South (CTA-South) in Chile~\cite{CTApage}, and the planned Southern Wide-field Gamma ray Observatory (SWGO)~\cite{SWGOpage}, which is currently in the process of selecting a site for installation in Argentina, Chile or Peru.

We should also mention that in 2023 an agreement was signed admitting Brazil as an Associate Member State of CERN. In addition, 
the 2026 edition of the International Conference on High Energy Physics (ICHEP) will be held in Latin America for the first time.

\section{HEP in other developing or emerging regions}
The Himalayan republic of Nepal has two major research universities - Tribhuvan University (TU, estd. 1959) and Kathmandu University (KU, estd. 1991). Neither university has a research program in high energy physics (HEP). However, Nepal has witnessed a significant number of activities in the past 10 years, primarily in capacity building - training high school teachers and university students. 

As far as capacity building is concerned, many schools and workshops have been held in Nepal. In 1989 the first edition of \href{ https://www.bcvspin.org/} {BCVSPIN} school was hosted by TU. Ever since, several editions of BCVSPIN have been held in many different countries with the most recent edition being held at KU in 2023. Since 2013, private contacts between universities and high schools in Nepal with CERN and ICTP personnel have led to nearly annual workshops, schools, and public outreach activities. These activities have resulted in students and teachers from Nepal getting trained at CERN. As of 2023, about 12 students from Nepal have been trained at CERN’s summer student program.  Similarly, about 10 high school teachers have been trained at CERN’s teacher training program. The teachers, in turn, continue to inspire their students. In addition, several high-level dignitaries from Nepal have visited CERN between 2013 and 2023, including the President in 2017 and the Prime Minister in 2019. 

In 2017, the first South Asian HEP Instrumentation workshop was hosted by KU. Soon after the workshop, an international cooperation agreement was signed between \href{https://cerncourier.com/a/cern-strengthens-ties-with-south-asia/} {CERN and the Ministry of Education and Science, Nepal}. In 2018, CERN donated 200 high-performance servers to build \href{https://cerncourier.com/a/boosting-high-performance-computing-in-nepal/} {Nepal’s first supercomputer}, which is being utilized by researchers in various domains, but not high energy particle physics. At this stage, there are several Nepali scientists holding PhD in HEP, working in the U.S., and many students in Nepal interested in HEP. However, we is not aware of any high-level government or university plan to launch high energy particle physics research in Nepal.

\section{Conclusions}
\label{sec:conc}
We have provided a description of HEP activities and efforts in Africa, Latin America and other developing or emerging regions. Community-driven efforts for providing a landscape of existing groups working in different projects in the different regions, and their available infrastructure is fundamental in order to explore synergies, coordinate activities and inform funding agencies with recommendations. Africa is currently in the process of developing its strategy on fundamental and applied physics and  Latin America has already concluded its first strategic planning in HECAP. An update of the process has already started.


\bibliographystyle{elsarticle-num}
\bibliography{myreferences} 

\end{document}